# DDS: DPU-optimized Disaggregated Storage [Extended Report]

Qizhen Zhang
University of Toronto
qz@cs.toronto.edu

Philip A. Bernstein
Microsoft Research
philbe@microsoft.com

Badrish Chandramouli
Microsoft Research
badrishc@microsoft.com

Jiasheng Hu
University of Toronto
jasonhu@cs.toronto.edu

Yiming Zheng
University of Toronto
yiming.zheng@mail.utoronto.ca

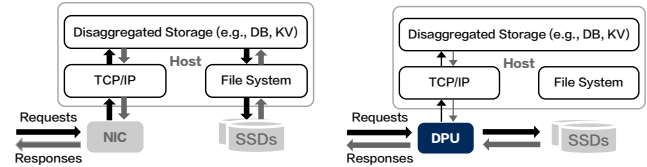

Figure 1: In current disaggregated storage (left), all storage requests are processed by the server host; with DDS (right), the majority can be offloaded to the DPU.

## ABSTRACT

The speed of storage devices and data center networks have significantly improved over the years. For cloud-native database systems (DBMSs), however, increasing bandwidth incurs higher latency and CPU consumption due to the disaggregation of storage. The overhead comes not only from the network and file modules of the OS kernel but also from the I/O stack inside the DBMS. Existing kernel-bypass techniques only partially reduce the latency and cost, and they require nontrivial DBMS modifications. This paper presents a new solution, DDS, a novel disaggregated storage architecture enabled by emerging networking hardware, namely DPUs (Data Processing Units). DPUs can optimize the latency and CPU consumption of disaggregated storage servers. However, utilizing DPUs for DBMSs requires careful design of the network and storage paths and the interface exposed to the DBMS. To fully benefit from DPUs, DDS heavily uses DMA, zero-copy, and userspace I/O to minimize overhead when improving throughput. It also introduces an offload engine that eliminates host CPUs by executing client requests directly on the DPU. Adopting DDS' API requires minimal DBMS modification. Our experimental study and production system integration show promising results. When saturating I/O performance, DDS achieves higher disaggregated storage throughput with an order of magnitude lower latency, and saves up to tens of CPU cores per storage server.

## 1 INTRODUCTION

A wide variety of data systems now run as cloud services, including database management systems (DBMSs), key-value (KV) stores, and document stores. A defining feature of cloud data systems is the disaggregation of storage and computation. In this architecture, application workloads execute on compute servers. Data is stored on dedicated storage servers that manage storage hardware and service storage requests. Storage I/O requests travel as messages over the data center network from compute servers to storage servers (see Figure 1 (left)). Decoupling compute and storage offers many benefits such as increased elasticity, durability, resource utilization, and hence reduced cost.

Storage disaggregation is used by cloud-native DBMSs, e.g., Aurora [65], SQL Hyperscale [15], PolarDB [19], and AlloyDB [30]. They store and replicate database files on storage-optimized servers[1] and execute read-only SQL queries and update transactions on compute-optimized servers. In most of these systems, only one compute server, called the *primary*, can execute update transactions. To scale compute, secondary compute servers can be added to execute read-only workloads; to scale storage, a large database can be partitioned across many storage nodes. This separation of compute and data allows DBMSs to host workloads of arbitrary scale, and allows users to pay only for what they need.

---
[1]In this paper, we refer to these servers as storage servers or storage nodes. Some service providers use different names, e.g., page servers in Azure SQL Hyperscale.

A major cost of these systems is the data communication needed to service reads and writes. The primary compute server ships its logs, rather than dirty pages, to secondary compute servers and storage servers; they replay the log records to refresh local pages. When a compute server has a cache miss, it reads the entire page from storage servers. Since pages are generally larger than log records and since reads outnumber writes in most workloads, page reads are a much larger fraction of data communication than log writes.

The CPU cost of servicing these page reads is large and growing. This is largely due to steadily increasing storage bandwidth, currently multiple GB/s for reads [13, 31, 33, 49]. Since the number of CPU instructions that DBMSs execute per page is fixed, increases in storage bandwidth lead to increases in CPU consumption. For instance, our benchmark of Windows Server 2022 shows that 2 GB/s disk I/O throughput (∼230K 8 KB page IOPS) consumes 5–6 dedicated state-of-the-art CPU cores. In the cloud, each core costs several thousand US$ over its server's lifetime [27].

The CPU cost of network protocols is also high. In the benchmark above, the CPU consumption for transferring pages at 2 GB/s with Windows Sockets on TCP/IP reached 14 cores at the sender.

In addition to CPU cost, the storage and network stacks increase request latency. Accessing a database page from locally attached SSDs typically takes 100–200 $\mu s$ [33], but read latency with disaggregated storage can be 10× worse. To avoid that high latency, some compute servers use local SSDs as an extended buffer pool, such as the resilient buffer pool extension in Azure SQL Hyperscale [15] and the block cache in AlloyDB [30]. Such caches increase service cost and the recovery time of compute server failures. Disaggregated KV services also suffer from high CPU cost and latency (§9).

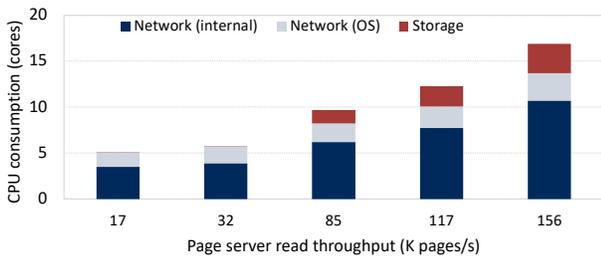

Figure 2: CPU cost of Hyperscale page server for reads.

Several technologies can move I/O into user space or hardware: Storage Performance Development Kit (SPDK) for storage access [1, 10]; and Data Plane Development Kit (DPDK) [62] and Remote Direct Memory Access (RDMA) for networking. These approaches eliminate OS overhead such as extra memory copies and context switches and can potentially reduce CPU cost and latency of disaggregated storage. However, they are incomplete solutions for two reasons: (1) to achieve high efficiency, they dedicate CPU cores to issue requests and poll completions, and thus still suffer from high CPU cost [22, 70]; and (2) they do not reduce the overhead of network and storage routines within data systems, which is higher than that of the OS as shown below.

To evaluate the potential benefit of kernel bypass, we measured SQL Hyperscale for a workload where a compute server randomly reads 8 KB pages from a page server that manages a 128 GB database (same hardware and OS setup as §8.1). Figure 2 shows the CPU consumption on the page server for different read throughputs. It also breaks out the network cost between the DBMS's internal network module and the OS network stack. The number of CPU cores consumed increases significantly with the target throughput, reaching 17 cores to read 156K pages/second. While all I/O components incur significant cost when the read throughput is high, the DBMS's network module contributes the most. This result motivates the need to reduce the I/O cost of disaggregation. *Kernel bypass (e.g., RDMA and DPDK) only partially eliminates the cost (the OS part) or requires redesigning the end-to-end communication stack, which incurs too much engineering expense and raises deployment issues for cloud DBMSs.*

In this paper, we present a new design to reduce the cost and latency of disaggregated storage for data systems. Our design uses a Data Processing Unit (DPU), a recent innovation in networking hardware. A DPU is a network interface card (NIC) equipped with compute and memory resources that are accessible by applications. It sits on the system bus, PCIe, which is the fastest path to access SSDs, and it directly interfaces to the network. It therefore can process storage requests as soon as they arrive at the NIC. Unlike SmartNICs based on FPGAs or ASICs, DPUs offer general-purpose compute resources and can thus support flexible storage offloading.

Using these insights, we propose DPU-optimized Disaggregated Storage (DDS). DDS optimizes disaggregated storage servers of cloud data systems such that *they consume minimal host resources to serve remote storage accesses with low latency.*

Figure 1 illustrates the differences between existing disaggregated storage and DDS. The former processes each request through multiple layers on the host: the OS to access the network and storage hardware via TCP/IP and the file system[2], and the data system to serve the request. By contrast, DDS directly accesses storage devices from the DPU.

One challenge in building DDS is choosing how to distribute functionality between the host server and its DPU. The host server has much more memory than the DPU and more powerful processors, which are needed for processing writes. The DPU has direct access to the network and storage devices and can execute reads with lower latency and lower cost than the host. Therefore, DDS offloads read requests to the DPU and processes update requests on the host.

To understand better this division of functionality, consider the processing of writes. In a cloud-native DBMS, writes are generated by sending the log to storage servers and secondary compute servers where they are replayed. For good response time to update transactions and to ensure later reads see fresh data, this must be done eagerly. Replaying updates requires complex logic and a large buffer cache, since many updates are applied to warm or hot pages. This is best done on the host. DPU memory is relatively small, so replaying updates on the DPU would generate many more SSD writes than on the host. Similar comments apply to a key-value store. Read-modify-write operations in a KV store often execute on warm data and thus benefit from running on the host, which can maintain a large buffer cache.

Now consider reads. A compute server typically has a large buffer cache, so it can offer good performance for queries over warm data. If a compute server asks a storage server to read a page, then the page is likely to be cold and not worth caching on either the host or DPU. Moreover, the logic to read a page is rather simple. Therefore, the DPU is well able to process the read, at lower latency and at lower cost since it avoids the overhead of DPU-to-host communication and is optimized for I/O efficiency.

Distributing storage functionality between host and DPU requires solving three problems: (1) how to enable read offloading to the DPU; (2) how to process update requests on the host but still minimize I/O overhead and keep the DPU updated with file mapping changes; and (3) how to handle cases where the data to be read is modified and cached on the host.

To solve (1), DDS provides a general and easy-to-use offload abstraction for users to supply customized code to the DPU to parse packets of interest into user requests and execute read requests.

For (2), DDS unifies host and DPU file operations with a file system that spans the DPU and host. A userspace library used by storage applications on the host serves as the front end to provide a familiar file API that minimizes application modification, and a file service running on the DPU serves as the back end to offload file execution and thus save host CPU cost. The DPU file service also ensures that the DPU sits on all I/O paths to disaggregated storage and hence has sufficient information for read offloading.

---

[2]Files are a common approach for data systems to manage data in secondary storage, e.g., DB pages and logs [15, 19] and KV records [20, 25] are often persisted in files.

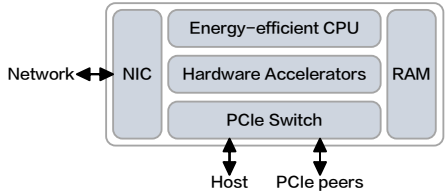

Figure 3: The general DPU architecture.

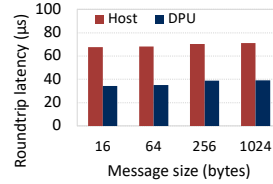

Figure 4: Responding to TCP messages on host vs. on DPU.

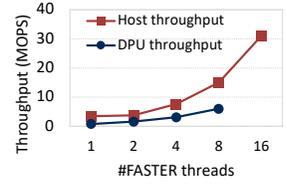

Figure 5: FASTER RMW throughput on host and DPU.

Problem (3) motivates *partial offloading*—read requests that access data cached on the host are better serviced by the host. Unfortunately, intercepting and processing some client-host communication in the DPU breaks end-to-end transport semantics between client and host. DDS introduces a network path that tackles the challenge and forwards read requests to the host when necessary.

We evaluated DDS in benchmark applications and two production data systems: a cloud DBMS and a key-value store service. DDS saves tens of CPU cores on each storage server in these systems. It also improves remote read latency up to an order of magnitude. Yet DDS requires minimal modification of the storage application, a crucial property for adoption. Incorporating DDS into the two production systems required only hundreds of lines of code changes.

This paper makes the following contributions.

- Characterizes the potential of DPUs for improving the performance and cost of disaggregated storage (§2).
- Presents a new architecture for reducing the cost and latency of host storage operations in a storage-disaggregated data system by offloading them to a DPU (§3).
- Presents the detailed design of a system, DDS, that realizes the architecture. It includes a unified storage path (§4), transparent network path (§5), and general offload engine (§6).
- Implements a prototype (§7), evaluates its efficiency (§8), and integrates it with two production systems (§9).

## 2 DATA PROCESSING UNITS

DPUs are SoC-based programmable NICs that are designed to offload host functionality. Compared to ASIC-only and FPGA-only SmartNICs [9, 23, 27], DPUs are easier to program, consume less energy, and achieve the same level of efficiency. Many chip companies offer DPU products, such as NVIDIA BlueField [8], Intel IPU [5], AMD Pensando [4] and Zynq MPSoCs [14], Broadcom Stingray [2], Kalray MPPA [6], and Marvell Octeon [7]. Cloud providers are also developing DPU chips, either in house (e.g., AWS Nitro [12] and Alibaba CIPU [11]) or via acquisition (e.g., Microsoft [3, 29]) or partnership (e.g., Google [56]). Many are already deployed at scale.

Figure 3 characterizes DPUs, which have five components:

- *High-speed network interface.* As networking devices, DPUs are designed to forward packets at high throughput in data centers. DPU NICs can provide hundreds of Gbps bandwidth.
- *Power-efficient CPU cores.* DPUs are equipped with CPUs to support general programming, e.g., control functions and generic user-space packet processing programs. DPU cores often adopt energy-efficient architectures. For instance, the CPUs in BlueField are low-power Arm cores.
- *On-board memory.* DPUs have memory to hold the working sets of the offloaded programs. Unlike GPUs and FPGAs that often integrate high bandwidth memory (HBM), DPUs mostly use lower cost DDR memory. As the interface and the on-board memory are often interconnected with on-chip networks, accessing DPU memory is faster than accessing host memory. The default memory size of current DPUs ranges between 16 GB and 32 GB with DDR4/5 DIMM extensions.
- *Built-in hardware accelerators.* DPUs harden compute-heavy data-path tasks, e.g., compression and regular expression matching, in on-board hardware accelerators. Executing corresponding workloads in hardware accelerators can be orders of magnitude faster than running them on CPUs.
- *PCIe support.* The final component is a PCIe switch that enables the DPU to access host resources, e.g., host memory via DMA. It also allows a DPU to directly access other PCIe devices, e.g., SSDs, via peer-to-peer PCIe communication.

**DPU opportunities.** These resources, if carefully designed, can be utilized to address the cost and performance issues in disaggregated storage. To reduce cost, DPUs employ techniques like userspace I/O and hardware offloading to highly optimize data-plane operations. Together with the general-purpose compute and memory resources, it is possible to build an ultra efficient engine on the DPU that offloads arbitrary I/O operations, thereby saving host CPU cycles.

To reduce latency, DPUs can cut the round-trips from the NIC to the host and bridge a fast path between a network request and the storage device. When a request arrives at the interface, rather than forward it to the host as today's NICs do, DPUs directly process it and access the data on SSDs. To demonstrate the benefits of doing so, we run a benchmark where a client sends TCP messages to a server, which echoes the messages back. The server has an NVIDIA BlueField-2 DPU. (See §8 for detailed configurations.) Figure 4 reports the message round-trip latency when the host server responds to the message vs. when the DPU directly responds, for different message sizes. We observe that *the DPU can halve the latency by avoiding forwarding the message to the host*.

**DPU limitations.** DPUs suffer from several constraints that prevent them from replacing the host servers in the design of cloud data systems. First, for energy efficiency DPUs have weaker and fewer cores than the host. For instance, in our testbed, the NVIDIA BlueField-2 (BF-2) DPU employs 8 Arm Cortex-A72 cores, which are much less powerful than the host's AMD EPYC processors. Since storage updates in cloud data systems do rather complex processing, offloading such workloads to DPUs can only decrease the performance. To quantify this effect, we evaluate the FASTER KV store [20] using the YCSB read-modify-write (RMW) benchmark on the host and on the BF-2 DPU. As shown in Figure 5, FASTER runs up to 4.5× slower on the DPU than on the host and can only scale to 8 threads. This shows that *executing update workloads on DPUs can considerably slow down data systems due to lower CPU performance*. In addition, DPUs often have insufficient memory for data systems to cache hot data on storage servers when serving large-scale workloads. In our scenario, the DPU has 16 GB on-board DDR4 memory, but production systems like Azure SQL Hyperscale and FASTER may allocate an order of magnitude larger memory to efficiently execute update workloads (e.g., log replay and RMW).

## 3 DDS OVERVIEW

**Design goals.** DDS seeks to bridge the gap between the resource requirements of data system storage servers and the characteristics of DPUs with *partial offloading*. Our specific goals follow.

1. *Minimal storage cost:* storage servers consume significant CPU capacity to support disaggregation. DDS seeks to reduce this CPU utilization by offloading much of it to the DPU.
2. *Minimal storage access latency:* the latency of reading from today's disaggregated storage is much higher than that of a raw SSD device. DDS seeks to close the gap.
3. *Ease of adoption:* data systems often require major changes to adopt new technologies. DDS seeks to enable data systems to obtain its benefits with minimal modification.
4. *Generality:* DDS seeks to offer mechanisms that are useful for a variety of cloud data systems.

We next present an architecture for achieving these goals (Figure 6). It consists of components that collectively optimize storage servers. We also outline research challenges in realizing the system.

**Architecture.** In DDS, a storage path unifies the application's file operations on the host and those that are offloaded to the DPU. Rather than process file reads and writes on the host's file system as in traditional storage servers, DDS moves file execution to a DPU *file service* and only leaves a lightweight *file library* for host applications to issue requests and poll for responses. As we showed in §1, existing host storage stacks are inefficient and CPU-costly. Executing file operations on the DPU saves these host CPU cycles.

As disaggregated storage requests are eventually converted to file operations for execution, unified file access is a prerequisite for DPU offloading. It allows DDS to maintain file metadata and the mapping of files to physical disk blocks (i.e., the *file mapping*) on the DPU. When a network request arrives from the network, the DPU constructs a file operation and executes it without consulting the host.

A network path then directs traffic between the host and the DPU to enable partial offloading. In traditional architectures, each client builds an end-to-end transport connection with the storage server. Network packets are processed by the TCP/IP stack in the host OS. By contrast, DDS separates network traffic at both the flow level and packet level. Within a flow, some packets are read requests that can be offloaded to the DPU while others should be forwarded to the host. At the packet level, DDS needs to divide batched requests into host and DPU subsets and direct them accordingly.

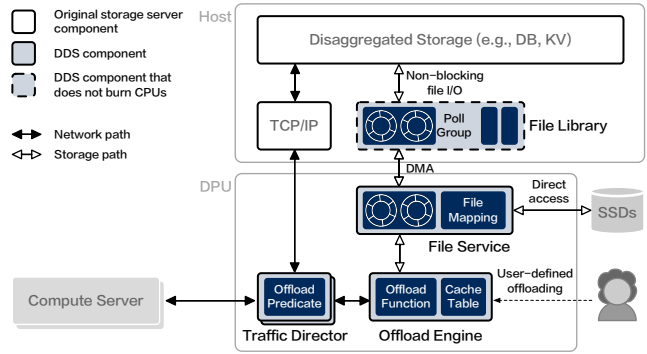

Figure 6: DDS architecture.

To examine every packet in a flow and every request in a packet, we develop a *traffic director*, a bump-in-the-wire running on DPU cores to process packets. It applies two pieces of user-provided information: *application signature*, which filters packets by their headers, and *offload predicate*, which inspects packet payloads.

Finally, an *offload engine* supports customizable offloading of storage functions to harness the capability enabled by the storage and network paths. Specifically, the offload engine on the DPU takes as input a remote storage request from the traffic director, outputs a file read operation, and directly submits it to the DPU file service. This process is guided by a user-supplied *offload function*. When data is read, it responds to the client via the traffic director.

**Challenges and key ideas.** DDS runs as an independent system that interacts with multiple layers, i.e., storage, network, and data systems. Working in harmony with existing systems while achieving all its goals has many challenges.

<u>File access efficiency and ease of use.</u> The separation of file front end and back end requires DDS to exchange data between the host and DPU. How can it do this with minimal performance degradataion? How does it execute file operations on the DPU using the wimpy cores? We solve these problems by using non-blocking, lock-free ring buffers, powered by DPU-issued DMA (§4.1), and by incorporating userspace I/O and zero-copy when accessing SSDs (§4.3). We carry out these ideas behind an interface that mimics today's file API to minimize application modification (§4.2).

<u>Offloading efficiency and generality.</u> One challenge in designing the offload engine is defining a general offload API. Unlike existing DPU SDKs (e.g., DOCA [53] and IPDK [34]) where users program low-level packets and flows, our API exposes objects data systems directly interact with. An obstacle to achieve generality is that request execution often needs application-specific state, such as the LSN in a storage page to serve a GetPage@LSN request [15]. We introduce an in-memory hash table on the DPU that users customize to cache information for translating user requests into file operations (§6.1).

Compared to a host file service, a challenge in achieving efficient execution of offloaded requests on the DPU is the asynchronous interfaces to both network and storage. We carefully allocate buffers and coordinate request execution to avoid any data copies (§6.2).

<u>Transport compatibility.</u> The bump-in-the-wire is first optimized with hardware acceleration (§5.1). More crucially, partial offloading breaks the end-to-end semantics of transport protocols, e.g., TCP. Offloaded packets will be considered lost by the host networking stack and thus unnecessarily trigger congestion control actions. To tackle this issue, we design the traffic director as a performance-enhancing proxy (PEP) that allows for packet redirection while still maintaining transport semantics. The PEP implements transport transparency to avoid redesigning the transport protocol and hence the host application that depends on that protocol (§5.2).

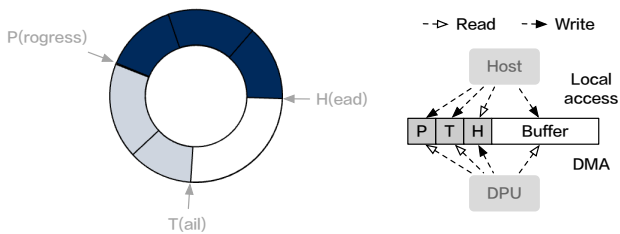

Figure 7: A progressive request ring: the logical view (left) and the physical layout and operations (right).

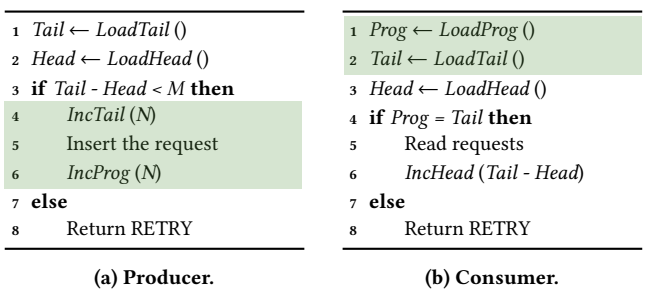

(a) Producer.   (b) Consumer.

Figure 8: Transferring requests from producers to the consumer (M = max allowable progress, N = request size). Highlighted are operations in critical order.

## 4 UNIFYING STORAGE

### 4.1 Message Movement

Moving file execution to the DPU requires file requests generated by host applications (i.e., reads and writes) to be transferred to the DPU, and responses to be transferred back after the execution. This host-DPU data movement is on the critical path of storage access, and thus we optimize it to achieve high overall performance (otherwise this could be a bottleneck, as §8.5 shows). At the center of our design are message buffers that support fast host-DPU communication. Storage applications on the host are normally multi-threaded. Each thread can access files. On the DPU, DDS dedicates one file service thread to retrieve requests and send back responses. This setup results in multi-producer single-consumer request buffers and single-producer multi-consumer response buffers. We focus the discussion primarily on request buffers.

In addition to performance, the first design goal of DDS poses two extra requirements on request buffers: (1) many threads on the host may concurrently issue file requests, but this contention should not burn CPU cycles; (2) communication can be CPU-expensive [22, 61], but host-DPU communication should not consume host resources. To that end, we introduce *DMA-backed lock-free ring buffers.* Figure 7 depicts the design. We explain first how producers and the consumer are coordinated with atomic operations and then how the host and DPU effectively utilize the ring with DMA.

Three pointers control the usage of the ring: the usual head and tail pointers of a ring buffer and *a new progress pointer* supporting concurrent insertions. The tail and progress pointers are atomically incremented by producers with compare-and-swap and read by the consumer with atomic load. The head is updated by the single consumer and read by producers. The ring's progress pointer is introduced to support multiple concurrent producers and to facilitate batching. Our proposal differs from existing ring buffer designs, e.g., Linux io_uring [43], in its ability to optimize host-DPU communication with DPU-issued DMA, e.g., opportunistic batching and reduced DMA operations. We discuss these benefits as follows.

Our buffer design creates a natural batching effect if multiple producers insert requests concurrently. We use a hyperparameter, maximum allowable progress, to control the batch size. To insert a request, a producer first checks the batch size (Lines 1–3 in Figure 8a). It returns RETRY if the size reaches the threshold M, indicating that insertions are outpacing consumption. Otherwise, the producer increments the tail to reserve space on the ring for the request, inserts the request, and then increments the progress pointer to indicate completion (Lines 4–6 in Figure 8a). To consume requests, the consumer first checks if *progress* and *tail* are equal (Lines 1–4 in Figure 8b), to see if it is safe to read requests. If so, it reads the batched requests and increments *head* (Lines 5–6 in Figure 8b). If not, some producer(s) have reserved space but not finished the insertion yet, so RETRY is returned. By using lock-free atomic instructions, we avoid spinlocks that burn CPU cycles or sleeping locks that sacrifice performance.

Figure 7 (right) shows the memory organization of the ring on the host, which we design to minimize host CPU consumption. First, the ring resides in the host memory such that *all the operations on the host are purely local-memory accesses, eliminating any CPU overhead associated with PCIe communication*, which is significant [22]. Second, the memory is pre-registered to the DPU driver so that the *DPU can directly access the ring via DMA without involving host CPUs*. In Figure 8, the only operations on the ring by the DPU are DMA-reads and DMA-writes. The ring memory consists of the pointer area that holds the three pointers of the ring and the data buffer where requests are inserted. The pointers are cache-line-aligned, to ensure that two threads accessing different pointers do not falsely contend (aka *false sharing*, where pointers are co-located on the same cache line). The physical ordering of these pointers is also optimized to reflect the order of operations, i.e., the progress pointer precedes the tail pointer. This enables a single DMA operation to read both P and T to perform the highlighted condition check in Figure 8b and is thus more efficient. (Placing T before P incurs two DMA reads: first P, then T.)

Response rings are similarly designed: the DPU is the single producer, and the host application threads are the consumers. §8.5 evaluates the performance of our DMA ring buffers.

### 4.2 Host Design

**File interface.** DDS' file library offers a familiar API to support common operations on files, directories, and notification groups. An application can invoke `CreateDirectory` to make a new directory, and put a file in the directory with `CreateFile`, which returns a file handle. Optionally, it can create an epoll-like notification group with `CreatePoll` and add the file to the group with `PollAdd`. To perform read and write operations on the file, it calls `ReadFile` and `WriteFile`. In addition to single reads and writes, the library supports gathered writes and scattered reads that take an array of source/destination buffers and perform one file I/O. The application can use the `PollWait` function to poll completions in a notification group, which waits a specified time for a response message before timeout. Except for `PollWait`, which we discuss shortly, all the operations are non-blocking to minimize the time and thus CPU cycles spent in the library.

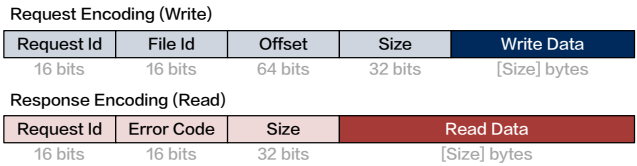

Figure 9: Encoding requests and responses on rings.

File operations can be classified into *data-plane* operations, which refer to the operations for file I/O, and *control-plane* operations, e.g., file and directory management functions. We focus our optimizations and discussion on data-plane operations (i.e., file reads and writes), which are the most important ones for the efficiency of a storage engine.

**Issuing requests.** The library manages file operations and completions with notification groups. When a notification group is created, the library allocates request and response ring buffers for the group and pre-registers them to the DPU driver for DMA. When the application performs a file operation, the library looks up the file's notification group, book-keeps this operation in the group, assigns it a request id, and inserts the request into the request ring. Figure 9 (top) shows the encoding of a file write on the request ring. In addition to the header that describes the request, the data to write is inlined in the request so that the entire request can be transferred to the DPU with a single DMA-read.

**Polling responses.** Polling file I/O completions in a notification group involves reading from its response ring. DDS supports two modes to handle the case where there is no response in the response ring when `PollWait` is invoked, *non-blocking* and *sleeping* (implemented with DPU driver interrupts), to accommodate various asynchronous I/O implementations. In the former mode the specified wait time is zero; the polling function returns immediately so that the caller thread can continue other compute tasks (or yield itself). In the latter mode, where the wait time is greater than zero, the caller thread will sleep. To wake up the sleeping thread when a response arrives, the DPU driver generates an interrupt when the response is DMA-written to the response ring. Both modes can achieve CPU-efficient polling—no CPU cycles are incurred in the sleeping mode, and the non-blocking mode gives the control to the application. It is up to the application to decide which approach is the easier adoption.

Figure 9 (bottom) describes how a read response is encoded on a response ring, consisting of the header and the read data. Like read requests, write responses have headers only. Once the file library polls a response, it uses the request id to locate the operation in its book-kept list and performs post-completion processing.

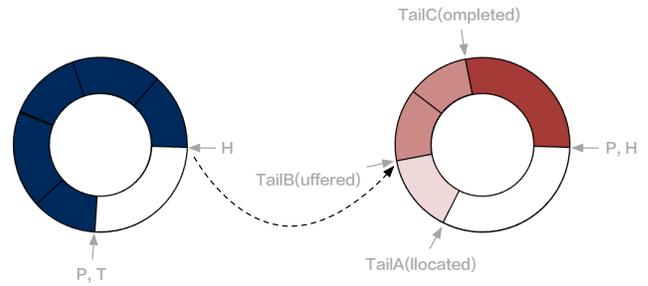

Figure 10: DPU executes requests (left) with zero-copy by pre-allocating the response (right).

### 4.3 DPU Design

For each notification group, the file service on the DPU maintains two buffers to synchronize messages with the request and response rings on the host (i.e., destination/source of DMA-reads/DMA-writes, respectively). A thread is dedicated to perform DMA to fetch requests and deliver responses via the algorithms in Figure 8.

**Low-latency file access.** Userspace storage I/O on DPUs allows user programs to bypass the OS overhead. For instance, NVIDIA BlueField includes SPDK in its storage drivers [10], and Intel IPU offers the infrastructure programmer development kit (IPDK) [34]. Hence we manage the PCIe-connected SSDs via these DPU-provided libraries to implement files and execute I/O.

Briefly, we divide and allocate SSD space with fixed-length segments (aligned by the disk block size), use a bitmap to track their availability, allocate disk space to files by segments, and group files with flat directories. One of the segments is reserved to persistently store the metadata of directories and files, as well as the file mapping, i.e., the vector of segments allocated to each file. Despite its simplicity, our file implementation suffices to support production systems as §9 shows. We leave evolving DDS files to a full-fledged DPU-based file system to future work (e.g., with more advanced space management, directories, and caching).

To execute a file I/O, the file service translates the file address into a disk block address using the file mapping and then submits the corresponding I/O operation to the userspace storage driver. When the I/O operation is completed, the file service generates a response incorporating the result of the execution and inserts it into the response buffer for delivery.

**Eliminating data copies.** File I/O is asynchronous in nature. If not treated with care, extra data copies may happen when executing requests and delivering responses. E.g., if the request buffer for receiving the DMA-read result from the host is shared between adjacent DMA-reads, the former request data must be copied before fetching new requests to avoid overwrites, so the DPU storage driver can consume it safely. The same issue applies to the response buffer and DMA-writes.

To circumvent the potential inefficiencies, our file service executes as follows. It first sets the size of the request buffer on the DPU to be the same as or greater than the request ring size on the host, so that *no outstanding requests overlap on the buffer*. This condition allows us to point the storage driver directly to the data in the request buffer as the input of the submitted I/O operations, hence eliminating the need for request copies. Moreover, before submitting an I/O operation, we reserve space in the response buffer in advance for the expected response and use the address as the destination of the I/O operation output, as Figure 10 shows. This is possible because the response header is well-defined in Figure 9, and for read requests we use the requested size as the read data size. Pre-allocation effectively eliminates the need for response copies.

**Ordered execution.** To preserve the request order in response space pre-allocation, we add two tail pointers in the response buffer. The original tail pointer, denoted by *TailC(ompleted)*, points to the end of the responses that have been delivered to the host response ring. The second tail pointer, *TailB(uffered)*, points to the end of the responses that have been finished by the file service but have yet to be delivered to the host. The final tail pointer, *TailA(llocated)*, represents the end of the pre-allocated response spaces. Initially, *TailA* and *TailB* are the same as *TailC*.

For each new request, the file service calculates its expected response size and advances *TailA* correspondingly. The status of the pre-allocated response (the error code field) is set as *pending*. The file service periodically checks the status of the pre-allocated responses, which on I/O completion is asynchronously changed to *successful* if the I/O succeeded or to an error code if it failed. Then, the file service advances *TailB* until a pending response. If *TailB* − *TailC* reaches a pre-configured batch size, then a DMA-write is issued to deliver the responses back to the host. Upon the completion of the DMA-write, *TailC* is advanced.

## 5 DIRECTING TRAFFIC

In DDS, disaggregated storage requests diverge at the DPU traffic director. We explain the design of this component, a key challenge raised by partial offloading, and how we address performance issues.

### 5.1 Bump-in-the-Wire

DPUs are designed as packet processing engines and generally support userspace storage I/O, e.g., DPDK [16, 34], and NIC-accelerated match-action rules [48]. The traffic director is built atop these capabilities as a bump-in-the-wire DPU application. It runs on the DPU CPU cores to inspect each packet arriving at the ingress port. Specifically, the inspection of a packet consists of two stages. First, the user-defined application signature in the packet header is applied to recognize if the packet belongs to the flows of interest. A signature specifies the 5-tuple of flows, i.e., client and server IP's and ports and the transport protocol. Below we show such an example, which matches any remote host (arbitrary IP address and port) as a client, local host with a specific port as the server, and TCP as the transport protocol:

$$[*:*, 10.10.1.1:1111, TCP]$$

As we can see, this stage only involves the packet's L3 and L4 headers. If a packet is matched, the traffic director proceeds to the next stage of executing further executes the user-defined offload predicate. Otherwise, the packet is directed to the host.

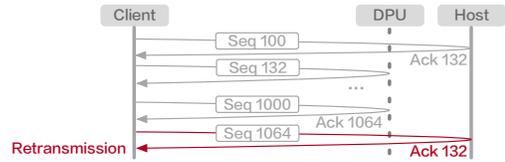

**Figure 11: Incompatibility between DPU partial offloading and reliable transport protocols, e.g., TCP.**

The second stage evaluates the offload predicate. The traffic director extracts the payload from a packet as *user messages* and inputs them to the offload predicate. For each user message, the offload predicate potentially outputs two messages, one for the host and one for DDS' offload engine. If there is a message for the host, the traffic director generates new packets and directs them to the host. (Details are in §6.1.) A message for the DDS offload engine will be shared with the offload engine for processing (see §6.2).

### 5.2 Transport Transparency

In storage-disaggregated data systems, a client often establishes a reliable network connection with the server to transfer requests and responses. The underlying transport protocol handles packet losses and congestion and ensures ordered delivery of correct data.

In particular, the most popular protocol in data systems, TCP, uses sequence numbers to achieve reliability and perform congestion control. When the server receives a packet, TCP checks the sequence number of the packet. If the number is not as expected for reasons like packet losses, out-of-order delivery, and in our case DPU offloading, TCP fast recovery is triggered, and a duplicate ACK is sent to the client. As a result, *the client will resend all the packets between the expected sequence number and the one received by the server*. Figure 11 shows an example: the storage server host first processes the packet with sequence number 100 and acknowledges it to the client. The offload predicate determines that the subsequent packets are offloaded to the DPU until the packet with sequence number 1064. When the host TCP receives this packet, it duplicates the ACK of 132 due to the unexpected sequence number. The client will resend all the requests that have been offloaded to the DPU.

Modifying the transport protocol to allow partial offloading not only complicates the design, but also requires modifying the data system that relies on the protocol and locks it into its implementation. We overcome this challenge by turning the traffic director into a *performance-enhancing proxy (PEP)*. A PEP is an in-network agent that improves the end-to-end performance of the transport protocol with only link-local changes and no transport-layer modifications [18, 69]. We design DDS's traffic director as a PEP for TCP splitting [55]. It automatically splits a single client-server TCP connection into two TCP connections: one between the client and the DPU and the other between the DPU and the storage server host. TCP splitting supports partial offloading as follows. All requests from the client are received by the traffic director on the DPU using the first connection. If a request cannot be offloaded, the traffic director sends it to the host with the second connection.

### 5.3 Performance Optimization

Directing traffic on the DPU may incur performance overhead. First, the bump-in-the-wire design requires the processing of all incoming packets. For on-path DPUs where the traffic always flows through the CPU cores [47], this has low overhead. However, for off-path DPUs like BlueField-2 (BF-2), the CPUs are not hardwired on the data path. Packets in flows of no interest must be forwarded to the host. We measured about 6$\mu$s of latency on BF-2 to forward a packet via an Arm core to the host. Second, TCP splitting requires a TCP/IP stack on the DPU. The built-in network stack in the OS on the SoC, e.g., Linux, introduces significant latency due to the kernel overhead, which is further exacerbated by weaker DPU cores.

To tackle the overhead, we first push the evaluation of the application signature down to the network interface, which has hardware support for line-rate routing based on packet headers. Only packets that match the signature are forwarded to the traffic director for further offload predicate evaluation—others are directly forwarded to the host. This avoids adding any latency to packets in flows of no interest. For offload predicate evaluation, we adapt and optimize a userspace TCP stack in the traffic director (detailed in §7).

Despite these optimizations, if a request matches the application signature but fails the offload predicate (e.g., a write request), the round trip to the traffic director still adds latency (∼10$\mu$s on BF-2). Applications can work around this overhead by isolating requests in different TCP connections and only allowing the ones that can be offloaded in the application signature.x

## 6 OFFLOADING TO DPU

We first present the interface that DDS offers for users to customize storage offloading, and then explain the techniques designed for the DPU that constitute an efficient offload engine.

### 6.1 Offload API

The API of DDS for offloading allows data systems to express storage operations at a high level. It is optimized for simplicity and generality based on the two-step process of DPU offloading in DDS: for each request, the user (1) determines whether it can be offloaded and (2) if positive, specifies how it should be offloaded, i.e., generating a file operation. OffPred in Table 1 performs Step (1). It takes as input a network message and the *cache table* (explained shortly) and outputs two lists of requests, one to offload to the DPU (*DPUReqs*) and the other for the host (*HostReqs*). Only one list can be empty, indicating no requests for that destination. This design accommodates batching where a single network message consists of multiple I/O requests, a common optimization to increase disaggregated storage throughput. A simple OffPred example for unrelated writes and reads is to put the former in *HostReqs* and the latter in *DPUReqs*.

Step (2) requires more user specification. For example, in a KV service a GET request contains only the key of the object. The index that maps keys to record locations in files is maintained on the host. In a DBMS the GET request contains the page id, from which the page's file location is directly derived. Hence, we introduce *offload function*, OffFunc, to translate a read request into a file read operation (see Table 1). In the above KV example, an OffFunc implementation can query the cache table for the given key to generate the parameters of a file read: file id, offset, and number of bytes to read. OffFunc is an imperative function defined by data system operators or service providers to translate user requests into file operations; it is not for general programming, e.g., memory allocation and other syscalls are disallowed for fast execution.

| Function | Return Value | API |
|---|---:|---|
| Offload predicate | *HostReqs*, *DPUReqs* | OffPred(*Msg, CacheTable*) |
| Offload function | *ReadOp* | OffFunc(*Req, CacheTable*) |
| Cache-on-write | *Keys*, *CacheItems* | Cache(*WriteOp*) |
| Invalidate-on-read | *Keys* | Invalidate(*ReadOp*) |

Table 1: Functions that users define to customize offloading in DDS. In gray are nullable values.

The cache table in DDS is an in-memory hash table that allows users to cache arbitrary object keys and data to support their offloading plans. The table is meant to summarize data stored in disk files. Writes modify the data in files and thus are the appropriate occasion for caching. Hence, we propose a *cache-on-write* mechanism to populate the table when a file write arrives from the host. When the host reads objects from files, the user may want to delete the objects from the table because they may be modified on the host. Hence, remote read requests to access these objects should be serviced by the host. The user-defined *invalidate-on-read* function serves this purpose. This cache table and the file mapping in the file service (§4.3) together provide a flexible two-level mapping that translates user requests in various data systems to file addresses and then to physical disk blocks, thereby facilitating general offloading.

As Table 1 shows, for each file write, Cache returns a list of object keys and their cache items to insert into the cache table; for each read, Invalidate returns keys whose cache entries will be removed. An example of *cache-on-write* and *invalidate-on-read* is to cache the file id, offset, and size of every object in a file write and invalidate the cache for every object in a file read.

### 6.2 Execution Engine

The execution of application signature and offload predicate were discussed in §5. We now show how the offload engine executes the offloaded read requests and manages the cache table.

**Executing offloaded reads.** When receiving a remote read request, the offload engine applies OffFunc to generate a file read: (*ReadOp* {*FileId, Offset, Size*}). To execute it, memory buffers must be allocated as the destination of the read data. A straw-man solution is to allocate *ReadOp.Size* bytes of memory on the fly, pass it to the file service, and once the read is finished, pass the read buffer to the traffic director to generate the final response for the client. This approach suffers from two memory copies: data is first copied from the file service to the read buffer and then copied in the traffic director to generate network packets. To avoid this overhead, our offload engine pre-allocates memory that accommodates both the read buffer and the packet buffer to achieve zero memory copies. Specifically, as Figure 12 illustrates, the offload engine reserves a pool of DMA-accessible huge pages. To execute a file read, it prepares the read buffer based on the size of the read (❶). The pointer to the buffer is then passed, along with other read parameters, to the DPU file service to perform the operation (❷). As these memory addresses are DMA-accessible, any additional memory copies between the pre-allocated read buffers and the storage devices are eliminated. When the read completes, an *indirect* packet buffer is created for the final network communication. This buffer contains only the placeholders for L4–L2 protocol headers. The read buffer is referenced as the payload of the packet. If the read size exceeds the MTU size of the network interface, multiple packets are used to segment the data (❸). The traffic director finally populates the packet headers with proper values and sends them to the remote client (❹). Doing so removes data copies between the offload engine and the traffic director. Zero-copy benefits are evaluated in §8.

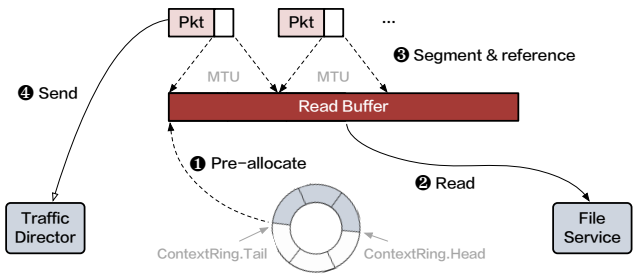

Figure 12: Executing offloaded read requests with zero-copy.

Since the DPU file service executes reads asynchronously, the offload engine needs to keep track of the outstanding operations and enforce ordering. To that end, it maintains a ring of contexts, each of which book-keeps the client id of the remote request, the metadata of the read operation, its completion status, and the pre-allocated read buffer. Asynchronous reads are coordinated using the context ring as follows, which is different from the execution of the file operations from the host (§4.3). For each offloaded read request, if the context ring is fully occupied, the offload engine sends it to the host via the traffic director. Otherwise, it utilizes the context at the current tail to perform the bookkeeping and sets the status of the read as pending. Then, it increments the tail and submits the read to the file service, which changes the status from pending to complete when the read completes. To process completions, the offload engine checks each pending read, starting from the current head: if the read is completed, it creates packets, sends them to the traffic director, and increments the head; otherwise, it stops to ensure ordering. The offload engine continually processes completions so that requests and responses can be executed in a timely manner.

Figure 13 shows how the offload engine execution coordinates asynchronous reads using the context ring. For each offloaded request, it first processes the completions of the previously issued file reads (Lines 3–4). If the context ring is fully occupied, the offload engine sends the current request and the remaining ones to the host (Lines 5–7). Otherwise, given the generated file read operation and the pre-allocated buffer, it utilizes the context at the current tail of the ring to perform the bookkeeping and sets the completion status of the request as pending (Lines 8–12). Then, it increments the tail of the context ring and submits the read operation to the file service (Lines 13–14). When the file service completes a read, it changes the status of the corresponding context from PENDING to COMPLETE. Hence, to process completions (Lines 18—27), the offload engine checks the status of each pending reads, starting from the current head: if the read is completed, it creates packets, sends them to the traffic director, and increments the head; otherwise, it immediately stops to ensure ordering. The offload engine constantly processes

```
1  while True do
2      Reqs ← receive new requests from the traffic director
3      for Req in Reqs do
4          CompletePending ()
5          if ContextRing.Tail cannot be incremented then
6              Send (remaining requests, host)
7              Break
8          ReadOp ← OffFunc (Req, CacheTable)
9          ReadBuf ← MemPool.Allocate (ReadOp.Size)
10         Ctxt ← ContextRing.At (ContextRing.Tail)
11         Bookkeep Req.ClientId, ReadOp, and ReadBuf in Ctxt
12         Ctxt.CompStatus ← PENDING
13         Increment ContextRing.Tail
14         SubmitToFileService (ReadOp, ReadBuf)
15     end
16     CompletePending ()
17 end

18 Function CompletePending ():
19     while ContextRing.Head ≠ ContextRing.Tail do
20         Ctxt ← ContextRing.At (ContextRing.Head)
21         if Ctxt.CompStatus = COMPLETE then
22             PktBufs ← CreatePkts (Ctxt.ReadBuf)
23             Send (PktBufs, Ctxt.ClientId)
24             Increment ContextRing.Head
25         else
26             Break
27     end
```

Figure 13: Offload engine execution.

completions such that requests and responses can be executed in a timely manner (Line 16).

**Managing the cache table.** The cache table is initialized by the offload engine and shared with the traffic director and the file service (see Table 2). When the file service executes a *host* file write/read, the user-provided *Cache*/*Invalidate* function is invoked to generate items for caching/invalidation. The file service then performs inserts or deletes to reflect the changes on the table. The traffic director and offload engine look up the table to execute OffPred and OffFunc respectively. The goal of the cache table is derived from this particular use scenario: the update rate in the file service is limited by the storage device performance, up to several millions of op/s, but the lookup throughput should not compromise DPU packet processing performance, which is up to tens of millions of packets/s. The cache table is hence optimized correspondingly. Specifically, we use cuckoo hashing to provide worst-case constant lookup time. We also chain items in a bucket to reduce the impact of collisions on insertions. As an additional optimization, DDS allows the user to specify the number of cache items allowable in the table, which is used to reserve the DPU memory for the table to avoid resizing the table at runtime.

**Comparison to existing work.** Compared to other SmartNIC offloading frameworks [37, 45, 52, 54], DDS offloads only storage requests that benefit from the DPU's capabilities. This partial offloading principle leads to a unique API design. Compared to offloading with computational storage [17, 40, 41, 46], DDS directly interacts with the network interface to service remote storage requests without involving the host. This host-bypass leads to higher gain from offloading but requires careful co-design of the network path (i.e., traffic director) and the offload engine. In general, the design of DDS' offload engine leverages the full programmability of the DPU SoC to serve diverse offloading needs of data systems.

| Component | Operation | Target Throughput |
|---|---|---|
| File Service | Insertion, Deletion | millions of op/s |
| Offload Engine | Lookup | millions of op/s |
| Traffic Director | Lookup | 10s millions of op/s |

Table 2: Uses of the cache table and their target performance.

## 7 IMPLEMENTATION

Our DPU platform is NVIDIA BlueField-2 (BF-2), which consists of a 100 Gbps network interface, 8 Armv8 A72 cores (64-bit, 80 KB, 1 MB, and 6 MB for L1, L2, and L3 cache respectively), 16 GB DDR4 memory, and data-path hardware accelerators, (e.g., a DEFLATE compressor and a RegEx accelerator), and is connected to the host server via PCIe Gen 4.0. We installed Ubuntu 20.04 as the DPU OS and DOCA [53] 1.5.1 as the DPU SDK. All DDS components together comprise 47,200 lines of C/C++ code [3].

**Host.** The DDS front-end library on the host is implemented with Microsoft Visual C++ and provides Windows file semantics. We separate the Windows shell from DDS host design (in §4.2) so other file interfaces, e.g., POSIX, can be layered atop DDS' front end.

**DPU.** We adopt SPDK for userspace storage I/O in the file service. After the DMA thread receives a file request, it invokes `spdk_thread_send_msg` to send the operation to a SPDK worker, which submits it with `spdk_bdev_read/write` to the NVMe SSD driver and, when the I/O completes, populates the response.

For the traffic director's userspace networking, we use Transport Layer Development Kit (TLDK) [28], a TCP/IP implementation based on DPDK. TLDK was built for Intel x86-64 processors, so we manually translate Intel SIMD intrinsics to Arm Neon intrinsics. We colocate the L2–L4 processing of each TCP connection on a single core to avoid inter-core coordination. Scaling up the traffic director to multiple Arm cores is realized using Receive Side Scaling (RSS), where we map TCP packets to DPU cores based on their 5-tuples. We carefully design the hash function for RSS to achieve symmetric TCP splitting, i.e., when the host responds in a connection, the response packets will be processed by the same DPU core that split the connection, which avoids sharing connection states between cores on the DPU.

**Resource utilization.** DDS minimizes resource consumption. On the DPU, unless otherwise noted, it utilizes three Arm cores of BF-2: one for DMA communication, one for the SPDK file service, and the remaining one for colocating the traffic director and the offload engine. Host CPU consumption is investigated in the next section.

---
[3]The prototype is available at https://github.com/microsoft/dds.

It also requires minimal memory footprint (lower than 1 GB on the DPU, excluding the cache table that is application-dependent).

## 8 EVALUATION

We evaluate DDS to answer the following questions:
- How many CPUs can DDS save on the storage server host?
- How much can DDS cut the latency of remote storage access?
- How effective are the optimizations in each component at improving the efficiency of DDS?

The first two questions correspond to the first two design goals of DDS. We investigate DDS's other goals in §9 with real case studies.

### 8.1 Methodology

The default cluster in our evaluation consists of two machines, each with two AMD EPYC 7325 24-core CPUs, 256 GB DDR4 memory, and a 1 TB NVMe SSD, and operated by Windows Server 2022 Datacenter. The storage server has a BF-2 DPU (§7), and the client connects to the server via an NVIDIA ConnectX-6 100 Gbps NIC.

We implement a storage-disaggregated application. The client issues random 1 KB file I/O requests. To stress storage server performance, it controls the request rate via parameters: the number of requests batched in a message, the number of outstanding messages, and the number of concurrent connections. The server receives requests, performs the file I/O, and returns the data to the client.

We measure the performance and cost of the storage server: overall throughput (rate of I/O request completions), end-to-end latency (time between issuing the request and receiving the response at the client), and the CPU consumption on the storage server host (number of CPU cores for executing the workload). We report incurred latency and CPU cost as a function of achieved throughput.

Our baseline does network I/O with Windows TCP sockets and file I/O with Windows NTFS. We also use these techniques in §9.

### 8.2 CPU Savings

We first investigate DDS's effectiveness in reducing CPUs consumed to perform I/O. For reads, applications benefit from DDS by performing file I/O with DDS's frontend library to replace the OS file system. Figure 14a shows this brings a noticeable CPU reduction—the baseline consumes 10.7 cores to achieve 390 K IOPS, while accessing files via the DDS library achieves 580 K IOPS with only 6.5 cores. To harvest the full benefit of DDS for reads, applications can completely execute requests on the DPU with DDS offload API. The figure shows that *DPU offloading effectively eliminates host CPU consumption* as DDS's efficient I/O stack on the DPU can drive read throughput up to 730 K IOPS. Achieving high-performance remote storage access with low host CPU cost only required implementing a 30-line `OffloadPred` and a 20-line `OffloadFunc`[4].

Write results are similar to those of reads, except that writes are slower (Figure 14b). DDS' offload API does not support writes, as updates in data systems are either compute-intensive or require more memory than a DPU supports (see §2). Still, executing writes with DDS frontend library saves more than 5 CPU cores compared to the baseline when write throughput exceeds 200 K IOPS.

---
[4]This application encodes file id, offset, and I/O size in the request, so `Cache` and `Invalidate` are not needed

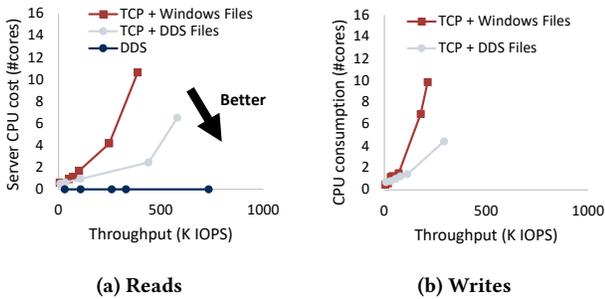

(a) Reads  (b) Writes

Figure 14: Achieved throughput vs. CPU consumption on the server (the number of CPU cores)

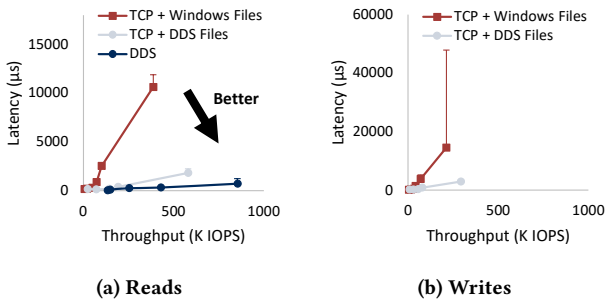

(a) Reads  (b) Writes

Figure 15: Achieved throughput vs. latency (dots and lines with a top show p50 and p99 respectively)

## 8.3 Latency Reduction

DDS can also improve remote storage access latency by removing OS file system overhead. Directly offloading reads to the DPU further decreases latency by avoiding the roundtrip to the host and its associated overhead. We now examine this benefit.

Figure 15a shows the results of the read benchmark. The baseline incurs the highest latency when achieving the same throughput due to the host I/O stack's overhead, as expected. Replacing the Windows file system by the DDS library yields a 6× latency reduction. *When all of the host overhead is bypassed with DDS offloading, the latency of read requests is improved by an order of magnitude.* Specifically, the latency of the baseline achieving 390 K IOPS is 11 *ms*, while DDS incurs only 780 *μs* when achieving 730 K IOPS.

Write latencies are shown in Figure 15b. The tail latency of the baseline for writes is surprisingly high, 48 *ms* when achieving 210 K IOPS. Replacing Windows files with DDS files is promising—lower tail latency (3 *ms*) when the throughput is higher (290 K IOPS).

## 8.4 Detailed Comparison

We dissect the benefits of DDS by comparing ten different storage solutions. To understand the overhead of storage disaggregation, we use Windows NTFS (❶) and DDS files (userspace front end on the host with file execution offloaded to the DPU ❷) to access SSDs locally[5]. To compare with application-implemented storage disaggregation, we adopt SMB [51] (❸), a remote file service in Windows

[5]We omit purely host-based userspace storage as Windows has limited support for DPDK, and our SPDK file service is not portable to the Windows host. Nevertheless, local storage access with DDS represents a stronger baseline because it exploits the SSD performance and avoids burning any host CPU cores.

Server that mounts remote disks to a local machine, and SMB Direct [50] (❹) that replaces TCP/IP in SMB with RDMA. To further isolate network and storage, in addition to TCP + Windows files (❺) and TCP + DDS files (❻), we replace TCP by Redy RPC [70]—an optimized RPC based on RDMA—to transfer storage requests over the network[6], i.e., Redy + Windows files (❼) and Redy + DDS files (❽). Finally, in addition to DDS's native design for TCP (❾), we ported RDMA to DDS to transfer messages between the client and DPU for file execution (❿). We evaluate these approaches using the read workload described in §8.1. We measure their peak storage throughput and the total CPU consumption on both the client and server (on a single machine in the case of local storage, i.e., ❶ and ❷) and end-to-end latency *when achieving the peak throughput*.

Figure 16 presents our results. With the traditional I/O stack (i.e., OS file system and TCP/IP), disaggregation degrades peak throughput and adds significant CPU and latency overhead (❺ vs. ❶). Although SMB Direct outperforms SMB due to the faster RDMA transport (❹ vs. ❸), both protocols have much lower throughput than application-controlled disaggregation (❸❹ vs. ❺-❿). The latter can use application optimizations such as batching to improve performance. When OS overhead is eliminated, disaggregated storage achieves the same peak throughput as local storage (❽-❿ vs. ❷). However, CPU consumption and latency are different across disaggregation solutions: while Redy with DDS files achieves low latency, some of its performance comes from burning a few CPU cores on both client and server; DDS (TCP) lowers host CPU consumption due to DPU offloading, but the TCP stack adds latency overhead. In comparison, DDS's (RDMA) numbers are close to those of local storage, incurring minimal disaggregation overhead.

RDMA is shown here only for performance and cost comparison. Using kernel-bypass techniques (e.g., DPDK and RDMA) to improve I/O efficiency in data systems requires careful and expensive stack redesign, as we discussed in Section 1. This experiment validates the benefits of DDS even when compared to these techniques.

## 8.5 Component Efficiency

We conduct microbenchmarks to evaluate each DDS component.

**Storage path efficiency.** Critical to DDS storage path performance is the DMA-based ring buffer that passes data between the host and DPU (§4.1). To evaluate its performance, we spawn host threads to send 8-byte messages to the DPU. We compare our progress-based lock-free proposal with two baselines: (1) FaRM-style ring buffer [26] where host threads set a flag after each message to indicate completions, and the DPU polls the completion and, after receiving a message, releases the space on the host ring buffer for future messages by clearing its bits, and (2) a lock-based ring buffer where host threads lock the ring before writing a message.

Figure 17a shows the message exchange rate with different numbers of producers. The FaRM-style ring has the lowest throughput (64 K OPS at peak) because it disables batching, and the DPU's polling overhead via PCIe is high. Batching with locks achieves high throughput (22 M op/s) when there is no concurrency on the host, but throughput drops to 1.4 M op/s with 64 producers due to host contention. Our proposal outperforms the baselines

[6]We use RDMA instead of TLDK on the host for the same reason above: TLDK is not supported by Windows. RDMA is also in general more performant and CPU-efficient.

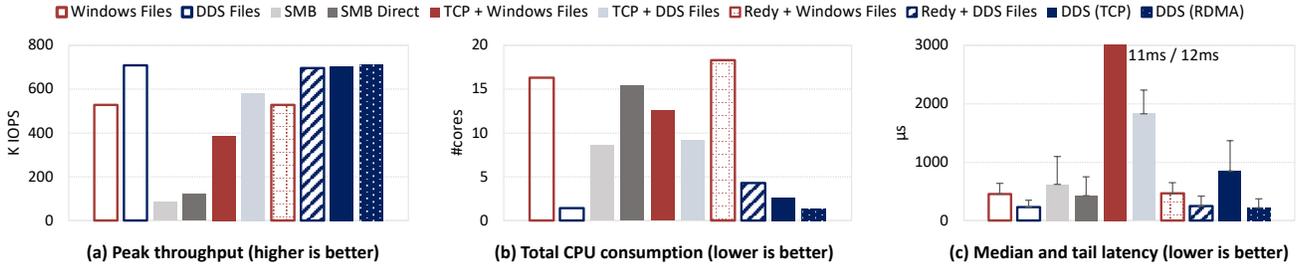

Figure 16: Comparing the peak throughput (a) of local storage (Windows files and DDS files) and disaggregated storage (SMB, SMB Direct, TCP + Windows files, TCP + DDS files, Redy + Windows files, Redy + DDS files, DDS offloading with TCP, and DDS offloading with RDMA), and the total CPU cost (b) and median and tail latency (c) when achieving the peak throughput.

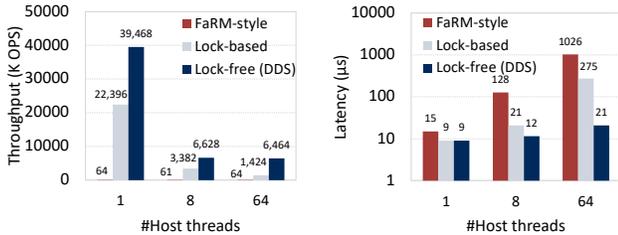

(a) Throughput  (b) Latency

Figure 17: DMA-based ring buffer performance

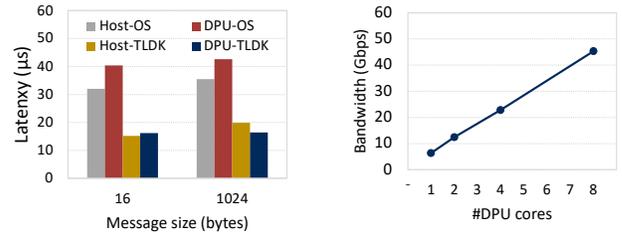

Figure 20: Comparing DPU offloading and kernel-bypass

Figure 21: The scalability of DDS traffic director

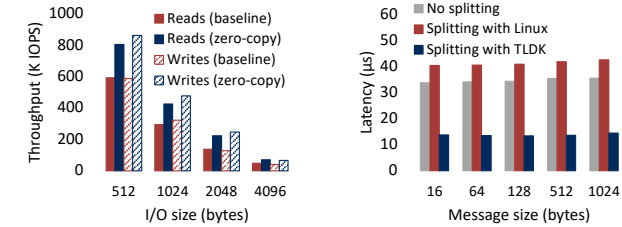

Figure 18: DPU-backed file I/O performance

Figure 19: Efficiency of TLDK vs. Linux for TCP splitting

significantly—it maintains 6.5 M op/s with 64 host threads, 10× and 4.5× higher than the FaRM-style and lock-based implementations respectively. Figure 17b shows the median latency of transferring individual messages. Compared to DDS' lock-free design, which achieves the lowest latency across all cases, the baselines show significant overhead: the FaRM-style approach uses many DPU-issued DMA reads for polling messages and an additional DMA write to release each message from the host, while lock contention between threads becomes the latency bottleneck in the lock-based approach when the number of concurrent producers increases.

We also evaluated the efficiency of DDS' storage path with host-issued file requests of different sizes (Figure 18). These results show that our zero-copy proposal in §4.3 is effective: compared to a baseline that pays extra memory copies to accommodate asynchronous I/O, DDS zero-copy design increases file throughput by up to 93%.

**Network path efficiency.** To showcase the benefit of userspace networking for DDS' traffic director (§5), we measure the server-side latency of an application where the client sends TCP messages to the server, which echos the same messages back (see Figure 19). We compare the vanilla version where the messages are echoed by the host versus the offloaded version where the echoing is done on the DPU. For the latter, we perform TCP splitting with two options: Linux TCP on BF-2 (OS) and TLDK (userspace). We have the following findings. First, Linux TCP introduces non-trivial overhead that offsets the benefit of DPU offloading. In fact, the latency of the offloaded echoing is even higher than that of the vanilla version. In comparison, our optimized use of TLDK avoids most of the cost and achieves 3× lower latency compared to Linux TCP, making DPU offloading beneficial (2.5× lower latency than the vanilla version).

To isolate the benefit of userspace networking from that of DPU offloading, we conduct an additional experiment where we run Linux (Ubuntu 20.04 with kernel 5.4.0) on the host to support TLDK and compare the performance of running TLDK on the host to that on the DPU for the application above. Figure 20 reports the result: processing large messages with TLDK on the DPU is faster (when the operation becomes more memory-intensive, which is the case in database systems, e.g., accessing pages). This is because (1) the round-trip from the NIC to the host is avoided, and (2) DPU memory is generally more efficient than host memory, as also observed in previous work [44, 63]. This experiment validates the significance of adopting userspace techniques to optimize DDS on the DPU.

The traffic director also provides high bandwidth. Figure 21 shows that it can direct 6.4 Gbps traffic with a single DPU core and, due to RSS, scale linearly when more cores are added.

**Offload engine efficiency.** Similar to the storage path, DDS' offload engine incorporates zero-copy to improve I/O performance (§6.2). To demonstrate its efficacy, we run the disaggregated application in §8.1 with and without this optimization. Figure 23 shows that avoiding memory copies on the data path improves both throughput (peak throughput increases from 520 K IOPS to 730 K IOPS) and latency (at peak throughput the latency drops from 250$\mu s$ to 170$\mu s$).

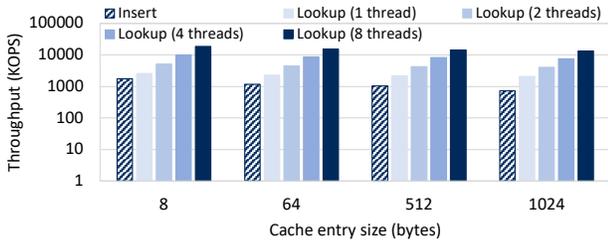

Figure 22: Cache table performance

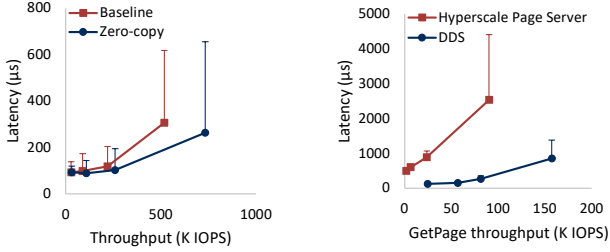

Figure 23: Impact of zero-copy on read latency

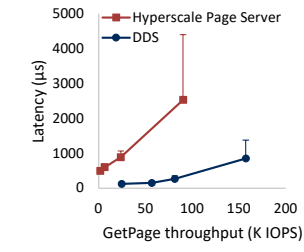

Figure 24: Throughput vs. latency of serving pages

We further investigate the performance of the cache table. We randomly generate cache items and measure the throughput of inserting them into and then looking them up from the cache table on the BF-2 DPU. For lookup, we vary the number of worker threads from 1 to 8. Figure 22 reports the results: on average, the cache table can achieve 1.2 M insertions/s with a single writer and 15.7 M lookups/s with eight readers across various cache item sizes. The requirements listed in Table 2 can thus be satisfied.

## 9 PRODUCTION SYSTEM INTEGRATION

To demonstrate the ease of use and performance of DDS for production data systems, we have integrated it with Azure SQL Hyperscale [15], the cloud-native version of Microsoft SQL Server, and FASTER [20], an open-source key-value store deployed at Microsoft.

### 9.1 Cloud DBMS

As a storage-disaggregated cloud DBMS, SQL Hyperscale uses *page servers* between the compute servers and the cloud storage service and adds log servers to further separate logging and data storage. A page server stores a partition of the database (e.g., 128 GB) on its local SSDs managed by the Resilient Buffer Pool Extension (RBPEX) and replays logs retrieved from the log servers to refresh the pages. On cache misses, compute servers send read requests to fetch data from page servers, which run a SQL stack atop Windows sockets and file system to service the requests. As Figure 2 shows, servicing read requests consumes significant CPUs on page servers. Hosting a large database means spawning a large number of page servers (e.g., a 100 TB database requires 800 page servers), which can easily become the dominant cost factor.

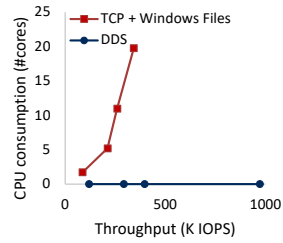

Figure 25: Disaggregated FASTER CPU cost (YCSB)

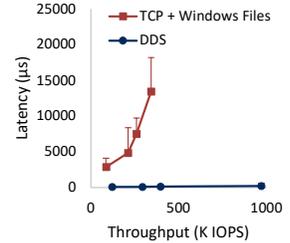

Figure 26: Disaggregated FASTER latency (YCSB)

Our SQL Hyperscale setup consists of a compute server, a page server, and a log server, each running on a machine specified in §8.1. The machine hosting the page server has a BF-2 DPU and integrates DDS as follows. First, we replace Windows NTFS with DDS front-end library to manage the RBPEX file. This needs only light-weight modification to the source code of SQL Hyperscale (hundreds of lines of code change, which is negligible compared to the size of the SQL Hyperscale repository), as our library provides the same file API as Windows files. Application-specific offloading also requires customizing the four functions in Table 1. Specifically, we need to cache the LSN and file offset of every page stored in the RBPEX file, keyed by page id (`Cache`) and invalidate it when the page server replays logs to update the page (`Invalidate`). When the traffic director detects a read request, it looks up the cache table using the page id and offloads the request if the cached LSN is equal to or greater than the requested LSN (`OffloadPred`). Finally, a read operation to the RBPEX file is constructed by `OffloadFunc`.

Offloading reads to the DPU effectively removes the CPU consumption in Figure 2. It also improves page serving latency, as Figure 24 shows: the page server incurs 4.4 *ms* at p99 to achieve 90 K IOPS, while 160 K IOPS only incurs 1.3 *ms* with DDS.

### 9.2 Key-value Service

FASTER stores KV records in a log that spans main memory and secondary storage. The in-memory section, i.e., the log tail, supports in-place updates. The rest is read-only and accessed via *IDevice*, the secondary storage abstraction. New records are appended to the tail and older records are flushed to IDevice if memory is insufficient. To execute a read, FASTER looks up the key in its hash index and retrieves the record from either memory or storage. When memory is constrained, most requests are serviced by IDevice. The default storage of FASTER is an IDevice implemented with Windows files.

We set up a disaggregated KV service with two machines of the same type as above, where the server runs FASTER with the YCSB benchmark [24] (8 B key and 8 B value) and stores most records in storage, and the client sends the YCSB uniform read workload to the server via TCP. Figures 25 and 26 show that the CPU consumption and latency are both high when increasing throughput: 340 K op/s costs 20 server cores and incurs 13 *ms* (18 *ms*) median (p99) latency.

To optimize the KV server with DDS, we first implement an IDevice with its front-end library. We then leverage its offload API in Table 1 to cache {key, file id, file offset, record size} entries in the cache table and use them to offload the workload to the DPU. All of these changes total 360 lines of code, and the improvement is significant. Figure 25 shows that FASTER with DDS achieves 970 K ops/s with zero host CPU investment. Benefiting from its efficient DPU I/O stack, DDS keeps latency as low as 300 $\mu s$.

## 10 RELATED WORK

Most research about DPUs has been done by the networking and distributed systems communities [27, 32, 35, 42, 45, 57, 60, 64, 66]. The systems community has explored DPU offloading of file system operations. We summarize some of these papers below.

IO-TCP [37] offloads media file serving to a DPU, but still forwards each request to the host. This would be inefficient for a data system due to the overhead of the OS and data system stacks. Xenic [59] is a main memory transaction system that uses a DPU to cache data. DDS can be used to cache data, but it can be used for many other scenarios too. Gimbal [52] leverages DPUs to improve channel utilization of remote storage in a multi-tenant setting, which we could use to extend DDS to better support multi-tenancy. hKVS[21] caches KV records on DPU memory and synchronizes writes between host and DPU. In contrast, DDS supports SSD-based data. LineFS [36] offloads CPU intensive activities in a distributed file system to the DPU. By contrast, DDS offloads remote read requests in a disaggregated storage setup. Lovelock [54] is a cluster design using DPUs with no hosts. It estimates cost and power of DPUs versus CPUs, but does not report on a prototype.

Offloading is also possible with recent storage techniques such as NVMe-oF, network-attached storage, and computational storage [17, 39–41, 46, 58], which offer applications direct access to the storage device. E.g., with firmware support, X-SSD [40] maintains a list of Logical Block Addresses to destage log data from persist memory directly to SSDs. In contrast, benefiting from DPUs' general programmability, DDS provides two-level address translation: the cache table that maps application requests to file addresses and the file mapping that maps file addresses to disk blocks. This flexibility allows customization for a variety of data systems, e.g., offloading operations on KV records or DBMS pages.

There is little exploration of using DPU-based SmartNICs by the database community. One exception is [63], which evaluated offloading two DBMS components to a DPU: a B-tree index and a global sequencer. Host throughput of the B-tree was greater than DPU throughput due to the wimpy DPU CPUs. For the sequencer, the DPU had higher throughput when using one-sided RDMA fetch-and-add operations. In both cases, the additional CPU capacity of the DPU increased overall throughput of the host+DPU system.

Pushdown is an effective query optimization strategy for disaggregated architectures [38, 67, 68, 71]. While query pushdown is challenging with DPUs due to their wimpy cores and limited memory, the hardware accelerators on DPUs can speed up certain database operators, e.g., the regular-expression ASIC can be leveraged for string operators. We leave this investigation to future work.

## 11 CONCLUSION

We showed that DPUs can significantly improve the performance and reduce the cost of disaggregated storage. We presented a architecture to realize these benefits, an implementation of the architecture, and an evaluation of its performance. Future work could explore offloading other DBMS functions to a DPU such as DBMS-specific networking and query operators, and utilizing DPU hardware accelerators such as encryption, compression, and regular-expression engines to execute compute-intensive components in cloud data system tasks in a portable manner.